\documentclass{webofc}
\usepackage[varg]{txfonts}   
\usepackage{srcltx} 
\def\1{\hbox{{1}\kern-.25em\hbox{l}}}
\begin{document}
\title{Improved calculation of the $\gamma^*\gamma \rightarrow \pi$
       process at low $Q^2$ \\
       using LCSR's and renormalization-group
       summation\thanks{This work is dedicated to the memory of Maxim Polyakov,
       a deeply admired colleague and friend of ours.}}
%
%

\author{\firstname{Sergey} \lastname{Mikhailov}\inst{1}\fnsep\thanks{\email{mikhs@theor.jinr.ru}} \and
        \firstname{Alexandr} \lastname{Pimikov}\inst{1}\fnsep\thanks{\email{pimikov@mail.ru}} \and
        \firstname{N.~G.} \lastname{Stefanis}\inst{2}\fnsep\thanks{\email{stefanis@tp2.ruhr-uni-bochum.de}}
}

\institute{Bogoliubov Laboratory of Theoretical Physics, JINR,
           141980 Dubna, Russia
\and
           Institut f\"{u}r Theoretische Physik II,
           Ruhr-Universit\"{a}t Bochum,
           D-44780 Bochum, Germany
          }

\abstract{
We study two versions of lightcone sum rules to calculate the
$\gamma^*\gamma\rightarrow\pi^0$ transition form factor (TFF) within QCD.
While the standard version is based on fixed-order perturbation theory
by means of a power-series expansion in the strong coupling, the
new method incorporates radiative corrections by renormalization-group
summation and generates an expansion within a generalized fractional analytic
perturbation theory involving only analytic couplings.
Using this scheme, we determine the relative nonperturbative parameters
and the first two Gegenbauer coefficients of the pion
distribution amplitude (DA) to obtain
TFF predictions in good agreement with the preliminary BESIII data, while the
best-fit pion DA satisfies the most recent lattice constraints on the
second moment of the pion DA at the three-loop level.
}
\maketitle
\section{Introduction}
\label{sec:intro}
In this paper we present our recent work on the calculation of the
two-photon process
$\gamma^*(Q^2)\gamma(q^2\sim 0)\rightarrow \pi^0$
which contains in the form of a convolution the distribution amplitude
of the pion \cite{Mikhailov:2021znq}---the simplest bound state in QCD.
Our analysis uses the method of lightcone sum rules (LCSR)'s
\cite{Balitsky:1989ry,Khodjamirian:1997tk} in the
extended form developed in \cite{Ayala:2018ifo}.
This scheme includes the QCD radiative corrections by means of
renormalization-group (RG) summation, ultimately amounting to a
generalized version of fractional analytic perturbation theory (FAPT),
invented in \cite{Bakulev:2005gw,Bakulev:2006ex} following
\cite{Karanikas:2001cs},
and reviewed in \cite{Bakulev:2008td,Stefanis:2009kv}.

We present the key elements of this formalism and apply it to the
recently released preliminary data of the BESIII Collaboration
\cite{Redmer:2018uew,Ablikim:2019hff},
which extend the range of measurements of the $\pi-\gamma$
transition form factor (TFF) to 
very low $Q^2\ll 1$~GeV$^2$
values with an unprecedented precision.
At such momenta the conventional LCSR method, based on fixed-order
perturbation theory (FOPT), cannot describe this transition process with
sufficient accuracy, though it works very well at high $Q^2$
\cite{Mikhailov:2016klg,Stefanis:2020rnd}.
Very recently, the full QCD calculation of the two-loop
coefficient function of leading twist
to the TFF $\gamma^*\gamma\rightarrow \pi^0$ was carried out analytically
by two different groups using different methods and obtaining coinciding
results \cite{Gao:2021iqq,Braun:2021grd}.
This level of computational accuracy of the NNLO radiative correction is
required by the expected precision of forthcoming data of the Belle II experiment.
In the higher $Q^2$ region, the measured values of $Q^2F(Q^2)$ reported by
the Belle Collaboration \cite{Uehara:2012ag} agree with the theoretical
expectations, while the previous \textit{BABAR} results \cite{Aubert:2009mc}
show above $Q^2\gtrsim 9$~GeV$^2$ a rapid growth with $Q^2$.
It is expected that Belle II will collect a large data sample with a
much higher accuracy to resolve this discrepancy allowing to test the
onset of factorization and identify the scale where asymptotic scaling
sets in \cite{Stefanis:2019cfn} or establish the existence of scaling
violations.

\begin{figure}[h]
\centering
\includegraphics[width=0.8\textwidth]{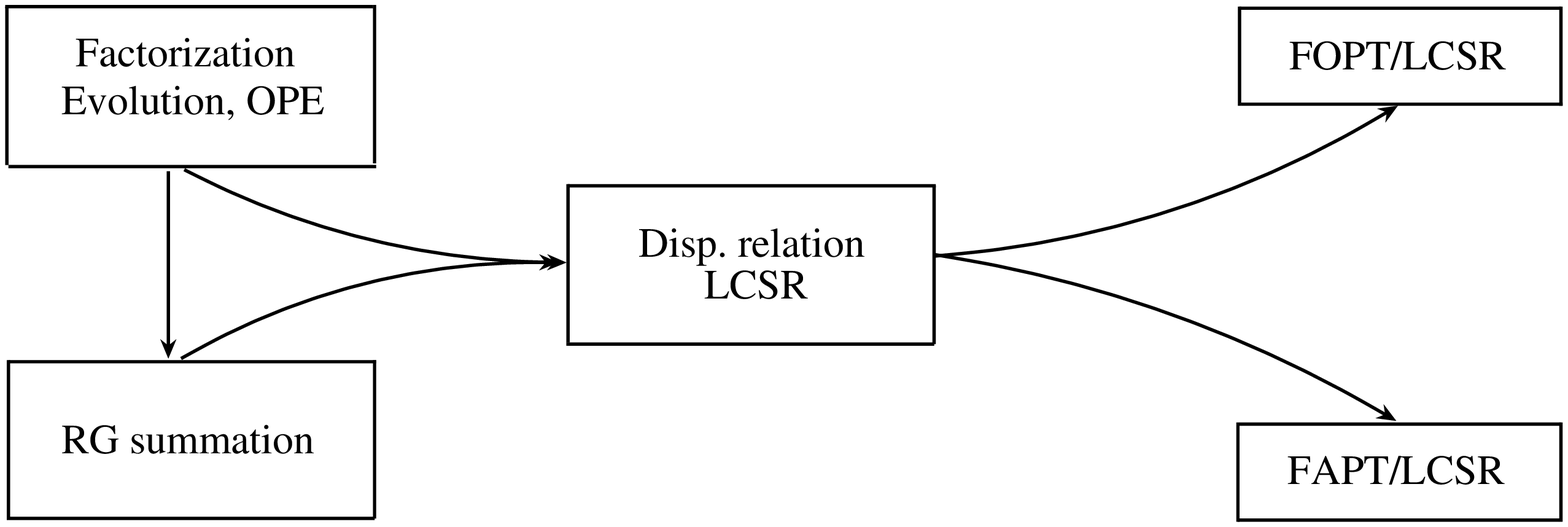} 
\caption{Schematic derivation of the FOPT and FAPT versions of LCSR's
to calculate the TFF.}
\label{fig:scheme}
\end{figure}

The main ingredients of both LCSR methods are shown in
Fig.\ \ref{fig:scheme}.

The paper is organized as follows.
In Sec.\ \ref{sec:FOPT-LCSR} we present a brief description of the
current status of the pion-photon transition form factor (TFF) calculation
using LCSR's in FOPT and point out its limitations at $Q^2\lesssim 1$~GeV$^2$.
In the subsequent section Sec.\ \ref{sec:FAPT-LCSR} we discuss the extended
version of the LCSR method which incorporates the summation of radiative
corrections via the renormalization group.
While the first version is sufficient to adequately describe this observable at
high-$Q^2$, see, \cite{Stefanis:2020rnd}, the second version is more suitable
at low momenta, where the radiative corrections, together with the higher
twists (twist-four and twist-six) contributions, are particularly important.

\section{Radiative corrections in FOPT/LCSR}
\label{sec:FOPT-LCSR}
In this section we briefly address the basic results for the TFF
using the method of LCSR's within FOPT in QCD.
Applying collinear factorization, the TFF for the hard exclusive process
$\gamma^*(Q^2)\gamma(q^2)\rightarrow\pi^0$
($Q^2\gg q^2\gg \Lambda_{\rm QCD}^2$) in leading twist two
can be written in convolution form as a power series expansion in the
strong coupling $a_s=\alpha_s/4\pi$ to get
\begin{equation}
F_\text{FOPT}^\text{(tw-2)}(Q^2,q^2)
			=
			N_\text{T}
			\left(
			T_{\rm LO}+ a_s T_{\rm NLO} + a_s^2 T_{\rm NNLO}+\ldots
			\right)
			\otimes
			\varphi_{\pi}^{(2)} \, ,
\label{eq:TFF-FOPT}
\end{equation}
where $N_\text{T}=\sqrt{2}f_\pi/3$.
The nonperturbative input of the pion structure is encoded in the
lightcone distribution amplitude (DA) of twist two (tw-2),
$\varphi_{\pi}^{(2)}$,
which describes the partition of the longitudinal momentum between
its two valence quarks with fractions
$x_q=x=(k^0+k^3)/(P^0+P^3)=k^+/P^+$
and
$x_{\bar{q}}=1-x\equiv \bar{x}$.
The hard coefficient functions are given by the following expressions
\cite{Mikhailov:2021znq}
\begin{subequations}
\label{eq:T}
\begin{eqnarray}
\!  T_{\rm LO}
& \! = \! &
  a_{s}^0(\mu_\text{F}^2)T_0(y)
\equiv
1/\left(
        q^2\bar{y} +Q^2y
  \right),
\\
\!  a_{s}T_{\rm NLO}
& \! = \! & a_{s}^1(\mu_\text{F}^2)T_0(y)
  \otimes \left[\mathcal{T}^{(1)}+
                      \underline{L~V_{0}}
              \right](y,x),
\label{eq:NLO}
\\
\!  a_{s}^2 T_{\rm NNLO}
& \! = \! &
  a_{s}^2(\mu_\text{F}^2)T_0(y)
  \otimes \!
  \left[\phantom{\underline{\underline{LV_1}}}\!\!\!\!\!\!\!\!\!
  \mathcal{T}_\beta^{(2)}
  - \underline{L \mathcal{T}^{(1)}\beta_0}
  + \underline{L \mathcal{T}^{(1)}\otimes V_0}
  -\underline{(L^2/2)\beta_0 V_0} \right.
\nonumber \\
&&  + \left. \underline{(L^2/2)V_0\otimes V_0}
  + \underline{\underline{LV_1}}
  \right](y,x)
  \, ,
\label{eq:hard-scat-series}
\end{eqnarray}
\end{subequations}
where
$L=L(y)=\ln[(q^2\bar{y}+Q^2y)/\mu_\text{F}^2]$
and using the following abbreviations: LO (leading order),
NLO (next-to-leading order) and NNLO (next-to-next-to-leading order).
Here the terms
${\bf \mathcal{T}^{(1)}, \mathcal{T}_\beta^{(2)}}$ and ${\bf \mathcal{T}^{(2)}}$
represent corrections due to parton subprocesses, while the
singly and doubly underlined terms are due to $\bar{a}_s(y)$ and
Efremov-Radyushkin-Brodsky-Lepage (ERBL) evolution \cite{Efremov:1978rn,Lepage:1980fj}
at one loop ($V_0$ kernel) and two loops ($V_1$ kernel), respectively.
On the other hand, the evolution of the pion DA is taken into account
in terms of the conformal expansion
\begin{equation}
  \varphi_{\pi}^{(2)}(z,\mu^2)
=
 \psi_0(x)
 + \sum_{n=2,4, \ldots}^{\infty}b_n(\mu^2) \psi_n(x)\, ,
\label{eq:gegenbauer}
\end{equation}
where
$\varphi_{\pi}^\text{asy}=\psi_{0}(x)=6x(1-x)\equiv 6x\bar{x}$
is the asymptotic pion DA.

\section{Radiative corrections using RG summation: FAPT/LCSR}
\label{sec:FAPT-LCSR}
To implement the RG summation in $F^\text{(tw-2)}$, we collect all
underlined evolution terms into the running coupling
$
 a_s(\mu^2)
\rightarrow
 \bar{a}_s(y)
\equiv
 \bar{a}_s\left(q^2\bar{y}+Q^2y\right)
$
and the ERBL factor \cite{Ayala:2018ifo} to obtain
\begin{eqnarray}\nonumber
F_{n}^\text{(tw-2)}(Q^2,q^2)\!\! &=& \!\!\!\!N_\text{T} T_0(y)
	\underset{y}{\otimes}
	\left\{
		\vphantom{\int_{a_s}^{\bar{a}_s(y)}}
		\!\left[\1
            +\bar{a}_s(y)\mathcal{T}^{(1)}(y,x)
            +\bar{a}_s^2(y)\mathcal{T}^{(2)}(y,x)
            + \ldots
		\right]
	\right. ~~~~
\\\label{eq:Tfin} \!\!\!\! &&  \left. \hspace{-20mm}
    \underset{x}{\otimes}
    \exp\left[-\int_{a_s}^{\bar{a}_s(y)}\!\!
		       \frac{V(\alpha;x,z)}{\beta(\alpha)} d\alpha
	    \right]
\right\}
                   \underset{z}\otimes \varphi_{\pi}^{(2)}(z,\mu^2) \, .
\end{eqnarray}
Employing in Eq.\ (\ref{eq:Tfin}) the Gegenbauer expansion given in Eq.\ (\ref{eq:gegenbauer}),
we get in leading logarithmic approximation (LLA)
\begin{eqnarray}
F_{n}^\text{(tw-2)}(Q^2,q^2)\stackrel{\text{1-loop}}{\longrightarrow} F_{(1l)n}^\text{(tw=2)}
=
  N_\text{T}T_0(y)
\label{eq:T1d}
  \underset{y}{\otimes}
  \left[\1+ \bar{a}_s(y)\mathcal{T}^{(1)}(y,x)\right]
   \left(\frac{\bar{a}_s(y)}{a_s(\mu^2)} \right)^{\nu_n}
   \underset{x}\otimes
   \psi_n(x) \, .
\end{eqnarray}
One realizes that for $q^2=0, ~ y\ll 1$, the summation of the
evolution terms via $\bar{a}_s(y)$ becomes inapplicable, even if
$Q^2$ is large \cite{Mikhailov:2021znq}.

We now show that this deficit is amended when a dispersion relation
for the TFF is involved.
The key elements of this procedure can be summarized as follows.
\begin{enumerate}
\item
Impose factorization and the twist expansion.
\item
Use a dispersive form of the TFF
\begin{eqnarray}
		\left[F_{}(Q^2,q^2)\right]_\text{an}
			= \int_{\!\!m^2}^{\infty}\!
				\frac{\rho_F(Q^2,s)}
				{s+q^2-i\epsilon}\,
				ds,~
		\rho_F(s)=\frac{\textbf{Im}}{\pi}\Big[F_{}(Q^2, -s)\Big] \label{eq:disper}
\end{eqnarray}
that inevitably leads to FAPT \cite{Bakulev:2005gw,Bakulev:2006ex}
with analytic couplings $\mathcal{A}_\nu$ (Euclidean space) and
$\mathfrak{A}_\nu$ (Minkowski space) in a nonpower series expansion.

\item
Dissect the spectral density using the perturbative expansion in
Eqs.\ (\ref{eq:Tfin}), (\ref{eq:T1d}), and employing $\rho_n$ for
each of the $\psi_n$ harmonics perform the twist expansion:
\begin{equation}
  \rho(Q^2,x)
=
  \sum_{0,2,4, \ldots} a_n(Q^2)\rho_n(Q^2,x)
  + \rho_\text{tw-4}(Q^{2}\!,x) +  \rho_\text{tw-6}(Q^{2}\!,x)
  + \ldots \, ,
\label{eq:spectral-density}
\end{equation}
where the integration variable in the spectral density
has been replaced by $s \to x=s/(Q^2+s)$ and $x_s=s_0/(Q^2+s_0)$.
Then, one obtains the LCSR based on Eq.\ (\ref{eq:disper}),
whereas the FAPT results are encapsulated in the analytic couplings
$\mathcal{A}_\nu, \mathfrak{A}_\nu, \mathcal{I}_\nu$.
The key point is that the imaginary parts in (\ref{eq:disper})
stem solely from the $a_s^\nu(y)$ factors in Eq.\ (\ref{eq:T1d}).
All details of this derivation are discussed in \cite{Mikhailov:2021znq}.
\item
To preserve the perturbative asymptotic limit of the TFF
$Q^2F(Q^2\rightarrow\infty)=\sqrt{2}f_\pi$, the calibration condition
$\mathcal{A}_\nu(0)=\mathfrak{A}_\nu(0)=0$ for $0<\nu\leq 1$ has to be
imposed \cite{Ayala:2018ifo}.
This yields generalized two-parameter FAPT couplings $\mathcal{I}_\nu$
with explicit expressions given in \cite{Mikhailov:2021znq}.
\end{enumerate}
Restricting for simplicity our attention to the NNLO$_\beta$ approximation
of the partial form factors $F_n$ within FAPT, we derive in the limits
$q^2\rightarrow 0, Q(y)\rightarrow yQ^2$ the following expression
\begin{eqnarray}
\!\!\!\! Q^2 F_\text{FAPT;n}^\text{(tw-2)}(Q^2)
\!\!\!\!&\thickapprox&\!\!\!\!
  \frac{N_\text{T}}{\underline{\left[a_s(\mu^2)\right]^{\nu_n}} \left[1+c_1 a_s(\mu^2)\right]^{\omega_n } }
                                       \bigg\{
\underline{\frac{\mathbb{A}_{\nu_n }(m^2,x)}{x}
                                       +\left(\frac{\mathbb{A}_{1+\nu_n}(m^2,y)}{y}\right)\underset{y}\otimes\mathcal{T}^{(1)}(y,x)} \nonumber \\
\!\!\!\!\!\!\!\!&& \hspace{-16mm} +\omega_n c_1\left[\frac{\mathbb{A}_{1+\nu_n}(m^2,x)}{x}+ \frac{\mathbb{A}_{2+\nu_n}(m^2,x)}{x}\frac{c_1(\omega_n-1)}{2}
+ \left(\frac{\mathbb{A}_{2+\nu_n}(m^2,y)}{y}\right)\underset{y}\otimes \mathcal{T}^{(1)}(y,x)\right] \nonumber \\
&& \hspace{-16mm}+ \underline{\underline{
\left(\frac{\mathbb{A}_{2+\nu_n}(m^2,y)}{y}\right)\underset{y}\otimes
\mathcal{T}^{(2)}(y,x)}} \bigg\}
                           \underset{x}\otimes
                    \psi_n(x) \, ,
\label{eq:NNLO}
\end{eqnarray}
where the terms contributing to the TFF in LLA are underlined.
The couplings $a_s^\nu(\mu^2) $ and
\begin{eqnarray}\nonumber
  \mathbb{A}_\nu(m^2,y)
=
  \theta\left(y\geqslant y_m\right) \left[\mathcal{A}_{\nu}(Q(y))-\mathfrak{A}_\nu(0)\right]
+ \theta\left(y < y_m\right)
 \left[\mathcal{I}_{\nu}(m(y),Q(y))-\mathfrak{A}_\nu(m(y))\right]
\label{eq:eff-coupl}
\end{eqnarray}
have to be evaluated with a two-loop running, while $c_1=\beta_1/\beta_0$
and $\omega_n= [\gamma_1(n)\beta_0-\gamma_0(n)\beta_1]/[2\beta_0\beta_1]$.
The only surviving term in the next-to-leading logarithmic approximation
(NLLA) for the numerically important case of the zero-harmonic
$(\omega_{n=0}=0)$ is the doubly underlined term.
For this reason, the effect of the two-loop evolution in the second
line is neglected, i.e., $c_1=0$
(for further details, see \cite{Mikhailov:2021znq}).

\section{FAPT/LCSR for the twist-4 pion TFF}
\label{sec:tw-4-TFF}
In the previous section we included the RG summation only in the
twist-two part of the TFF.
To carry out a comprehensive analysis of the experimental data in the
low-$Q^2$ regime $0.35\leq Q^2 \leq 3.1$~GeV$^2$, covered by the
CELLO \cite{Behrend:1990sr}, CLEO \cite{Gronberg:1997fj},
and the preliminary BESIII \cite{Redmer:2018uew,Ablikim:2019hff}
data, we have to extend the RG procedure to the twist-four term
stemming from the contribution of the two-particle DA.
To this end, we make the assumption that the three-particle DA
term is modified after the RG summation in the same way as the
twist-two part and evaluate the complete twist-four contribution
in an analogous way.

\begin{figure}[h]
\centering
\includegraphics[width=0.5\textwidth,clip]{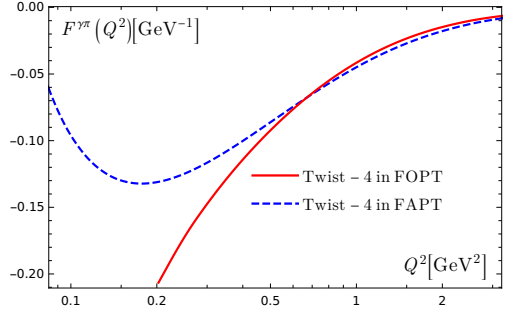}
\caption{Comparison of the twist-four contribution obtained with
FOPT/LCSR's (red solid line) and FAPT/LCSR's
(blue dashed line) using in Eq.\ (\ref{eq:AppB-HV}) the single
delta-function resonance model with $k(M^2)=1$.}
\label{fig:tw-4}
\end{figure}

Following \cite{Khodjamirian:1997tk}, we recast the TFF in the following
form
\begin{eqnarray}
	F^{\gamma\pi,\text{tw-4}}_{\text{FAPT}}\!\left(Q^2\right)
	\!=\!
	\frac{\sqrt{2}f_\pi}{3Q^2}\!\left[H_\text{FAPT}^\text{tw-4}(Q^2)\!+\!\frac{Q^2}{m^2_\rho} k(M^2)V_\text{FAPT}^\text{tw-4}(Q^2,M^2)\right],
\label{eq:AppB-HV}
\end{eqnarray}
\begin{eqnarray}
	V_\text{FAPT}^\text{tw-4}(Q^2)=-\frac{\delta^2_\text{tw-4}(\mu_0^2)}{M^2}
	\int_{\bar x_s}^{1} \!\!dx\, \frac{\Delta_\nu(s_0,\bar x)}{(a_s(\mu_0^2))^\nu}\frac{\varphi^{(4)}(\bar x)}{x^2}
	\exp\left(
	\frac{m_{\rho}^2-Q^2\bar x/x}{M^2}
	\right)\, ,
\end{eqnarray}
\begin{eqnarray}
	H_\text{FAPT}^\text{tw-4}(Q^2)=
	-
	\frac{\delta^2_\text{tw-4}(\mu_0^2)}{(a_s(\mu_0^2))^\nu} Q^2
	\mathbb A_\nu(s_0;u) \!\underset{u}\otimes \frac{\varphi^{(4)}(u)}{Q^2(u)}\, ,
\end{eqnarray}
where we use the same notations and definitions as in \cite{Mikhailov:2021znq}.
The expression for the function $\varphi^{(4)}(x)$ below has two terms:
the first term in the parenthesis is the two-particle DA contribution while
the second term is due to the three-particle DA:
\begin{eqnarray}
	\varphi^{(4)}(x)=\left(\frac{50}3+10 \right) x^2(1-x)^2\, .
\end{eqnarray}
Note that the part coming from the three-particle DA is included here only
heuristically.
Due to the RG summation, the three-particle contribution to twist-four should
be considered more rigorously, but this lies beyond the scope of this work.
For completion, we also quote the value of the twist-four coupling parameter $\delta^2$
together with its FOPT evolution,
\begin{eqnarray}
&&
	\delta^2_\text{tw-4}(\mu^2_0)= 0.95\, \lambda^2_q/2 = 0.19~\text{GeV}^2\,, \nonumber \\
	&&
	\delta^2_\text{tw-4}(Q^2)
	= \left[\frac{a_s(Q^2)}{a_s(\mu^2_0)}
	\right]^\nu\delta_\text{tw-4}^2(\mu^2_0)\,,
	\nu=\gamma_{T4}/\beta_0\, ,\gamma_{T4} = 32/9~\text{\cite{Bakulev:2002uc}}\,.
\label{eq:twist-4}
\end{eqnarray}
The results for $F^{\gamma\pi,\text{tw-4}}\left(Q^2\right)$
in FAPT compared to FOPT are displayed in Fig.\ \ref{fig:tw-4}.

One may conclude that the FAPT modification of the twist-four
contribution provides protection against drastic changes relative to
the FOPT behavior, extending this way the applicability domain of the
FAPT TFF calculation to lower $Q^2$ values.
On the other hand, both curves come close to each other at $Q^2 > 0.35$~GeV$^2$
so that the estimated ratio (twist-4)/(twist-2) becomes approximately $1/3$
near the lowest BESIII data point, see \cite{Redmer:2018uew,Ablikim:2019hff} and
Table I in \cite{Mikhailov:2021znq}.

\section{Phenomenological analysis}
\label{sec:results}
\begin{figure}[h]
	\centering
	\includegraphics[width=0.8\textwidth,clip]{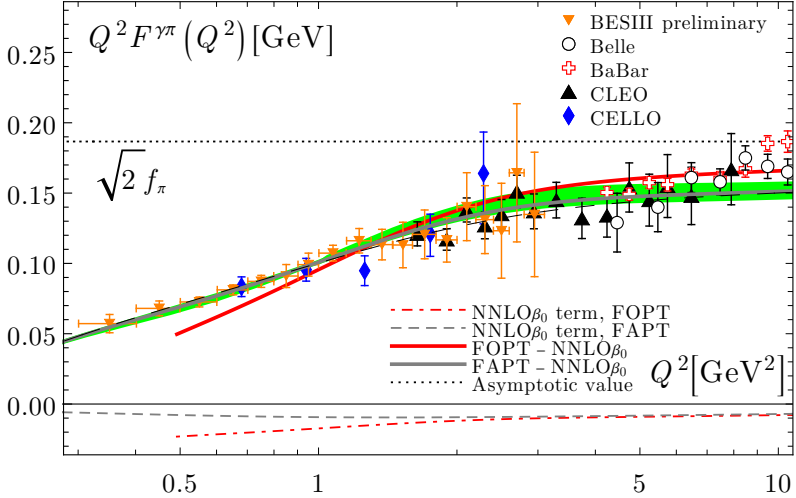}
	\caption{
This figure serves to demonstrate that the TFF calculated with FAPT/LCSR's
(using $N_f=3$) compares well with the data up to momenta where it is
expected to enter the scaling regime predicted by perturbative QCD, though with reduced
accuracy compared to the FOPT/LCSR result, see \cite{Stefanis:2020rnd}.
The displayed curves are explained in the text.}
\label{fig_TFF-predictions}
\end{figure}

In this section we present the pivotal phenomenological results of
our analysis based on FAPT/LCSR's.
This is done for simplicity in terms of
$F_\text{LCSR}^{\gamma\pi}(Q^2)$ for the
harmonics $\psi_n$, $F_\text{LCSR;n}^{\gamma\pi}$,
and predictions are shown in Fig.\ \ref{fig_TFF-predictions}
for some selected pion DA's including the best-fit one determined
in \cite{Mikhailov:2021znq}.
Various data from different experiments in the momentum range
$0.35\leq Q^2 \leq 10$~GeV$^2$ are also shown to effect the quality of the
fitting procedure at low $Q^2$.

To this end, we determine the following nonperturbative parameters:
\begin{enumerate}
\item
$b_2, b_4$ (conformal coefficients of the twist-2 pion DA)
using the following constraints at the scale $\mu_{0}^{2}=1$~GeV$^2$:
$b_2(\mu_{0}^{2})=[0.146,0.272]$,~ $b_4(\mu_{0}^{2})=[-0.23,-0.049]$
(domain of twist-2 BMS DA's) \cite{Bakulev:2001pa,Bakulev:2002uc}.
\item
$\delta_\text{tw-4}^2(\mu_{0}^{2})=0.19\pm 0.04$~GeV$^2$
(twist-4 coupling parameter) \cite{Bakulev:2002uc}.
\item
$\delta_\text{tw-6}^2(\mu_{0}^{2})=(1.61\pm 0.26)\times 10^{-4}$~GeV$^6$
(twist-6 coupling parameter)
extracted from the data in \cite{Mikhailov:2021znq}.
\end{enumerate}

A best fit to the $1\sigma$ and $2\sigma$ error ellipses of the data
up to 3.1~GeV$^2$ is given in \cite{Mikhailov:2021znq} with $\chi^2_\text{ndf}=0.38$.
It provides agreement with the N$^3$LO lattice constraints from
\cite{Bali:2019dqc} and yields $b_2(\mu_{0}^{2})=0.159$,
which corresponds to $b_4(\mu_{0}^{2})=-0.098$ if depicted within the BMS domain.
Note that the platykurtic (pk) range of pion DA's \cite{Stefanis:2014nla,Stefanis:2015qha}
lies entirely within the $1\sigma$ error ellipse of this data set
and is close to the NNLO lattice strip \cite{Bali:2019dqc}.

The TFF predictions calculated this way, are given by
\begin{equation}
  F^{\gamma\pi}_\text{LCSR}\left(Q^2\right)
=
  F^{\gamma\pi}_{\text{LCSR};0}\left(Q^2\right)
  + \sum_{n=2,4} b_n(\mu^2)F^{\gamma\pi}_{\text{LCSR};n}\left(Q^2\right)
  + \delta_\text{tw-4}^2(\mu^2) F^{\gamma\pi}_\text{tw-4}\left(Q^2\right)
  + \delta_\text{tw-6}^2(\mu^2) F^{\gamma\pi}_\text{tw-6}\left(Q^2\right)
\label{eq:TFF-predictions}
\end{equation}
and are displayed in Fig.\ \ref{fig_TFF-predictions} for the scaled TFF
in comparison with various data.

The green strip shows the theoretical uncertainties related to the
BMS DAs calculated with QCD sum rules with nonlocal condensates
\cite{Bakulev:2001pa}.
The induced uncertainty in the low-$Q^2$ tail is comparable with the
errors of the BESIII experiment, while at higher momenta it is smaller
than the data errors.
The best-fit results, obtained with the pion DA determined in \cite{Mikhailov:2021znq},
are denoted by the grey solid line (FAPT/LCSR)
and the red solid line (FOPT/LCSR in NNLO), respectively.
One notices that the analogous prediction obtained with the platykurtic
pion DA \cite{Stefanis:2014nla} within FAPT/LCSR (black dashed line)
coincides with the outcome based on the best-fit at the level of $\chi^2_\text{ndf}=0.57$.
This is significant because this DA amalgamates the key features of NLC's
parameterized by the vacuum quark virtuality
$\lambda_{q}^{2}=0.45$~GeV$^2$ (entailing endpoint suppression) with
unimodality---in contrast to the bimodal DA's in the BMS domain with
$\lambda_{q}^{2}=0.40$~GeV$^2$.
We remind that unimodality is welcome because this is a prominent
characteristic of pion DA's induced by dynamical chiral symmetry breaking
and the emergent generation of hadron mass, see for a recent review
\cite{Roberts:2021nhw}.
However, such DA's have enhanced tails and fail to reproduce the data
with a good accuracy \cite{Stefanis:2020rnd}.

Finally we note that the dashed curves at the bottom of
Fig.\ \ref{fig_TFF-predictions} representing the dominating contribution
NNLO$_{\beta}$ in FOPT (dashed-dotted line) and FAPT (dashed line)
yield above $Q^2>2$~GeV$^2$ comparable results.
This makes it apparent that below $Q^2\lesssim 1$~GeV$^2$, the RG summation
of the radiative corrections prevents the overestimation of the NNLO
contribution in the FOPT/LCSR scheme.
A detailed comparison with the total NNLO contribution
\cite{Gao:2021iqq,Braun:2021grd}
within our scheme would be useful.

\section{Conclusions}
\label{sec:concl}
In this work we have given a brief presentation of our recent detailed
analysis in \cite{Mikhailov:2021znq} dealing with the extension of the
method of LCSR's provided by the implementation of RG summation of QCD
radiative corrections to the $\gamma^*\gamma\rightarrow\pi^0$ TFF.
We showed that this new scheme inevitably leads to a different perturbative
expansion.
Instead of a power series expansion in terms of the strong coupling,
as in FOPT, on which the standard LCSR method is based, the new perturbation
theory employs only analytic couplings amounting to an improved version of
FAPT \cite{Ayala:2018ifo}.
We used the new FAPT/LCSR scheme \cite{Mikhailov:2021znq} to calculate the
$F^{\gamma\pi}(Q^2)$ TFF and determined the  employed nonperturbative parameters
$b_2, b_4, \delta_\text{tw-4}^2, \delta_\text{tw-6}^2$.
To this end, we employed the new preliminary BESIII data
\cite{Redmer:2018uew,Ablikim:2019hff} in combination with the most advanced
lattice calculation of the second moment of the pion DA in three loops
\cite{Bali:2019dqc} and found a bimodal DA which provides a good compromise
for all these constraints.
The obtained TFF predictions presented above agree with reasonable accuracy
also with other data up to $Q^2=5$~GeV$^2$ and beyond with an accuracy that
supersedes that of calculations within FOPT at $Q^2<1$~GeV$^2$.
Remarkably, the TFF predictions in the FAPT/LCSR scheme obtained with
the new best-fit pion DA (Fig.\ \ref{fig_TFF-predictions}) almost coincide
with those calculated with the platykurtic pion DA \cite{Stefanis:2014nla},
while at the same time both of them are within the uncertainty range
(shaded green strip) determined by QCD sum rules with nonlocal condensates
\cite{Bakulev:2001pa}.
This is important because this DA has a unimodal profile that finds
support by the findings of other investigations
\cite{Zhang:2017bzy,Ji:2020ect,Roberts:2021nhw} which, however, cannot reproduce the
Belle data \cite{Uehara:2012ag} with the same high accuracy because
in contrast to the pk DA, the derived pion DAs are endpoint enhanced
\cite{Stefanis:2020rnd}.
The expected high-precision measurements by the Belle-II Collaboration may
provide adjudicative constraints to select the most appropriate pion DA.

\section*{Acknowledgments}
S.~V.~M. acknowledges support from the Heisenberg-Landau Program 2021.


\end{document}